\documentstyle[a4,12pt]{article}
\begin{document}
\parskip=5pt plus 1pt minus 1pt

\begin{flushright}
{\large\bf LMU 01/99}
\end{flushright}
\begin{flushright}
{\large\bf January 1999}
\end{flushright}

\vspace{0.2cm}

\begin{center}
{\large\bf The Production of Single $t$--Quarks at LEP and HERA}
\end{center}

\vspace{0.3cm}

\begin{center}
{\bf Harald Fritzsch} \footnote{Electronic address:
bm@hep.physik.uni-muenchen.de} \\
{\bf and}\\
{\bf Dirk Holtmannsp\"otter}\\
\bigskip
{\it Sektion Physik, Universit$\ddot{a}$t M$\ddot{u}$nchen,
Theresienstra$\beta$e 37, D-80333 M$\ddot{u}$nchen, Germany}
\end{center}

\vspace{1.8cm}

\begin{abstract}
We study the possibility to produce single $t$--quarks both at LEP II and
HERA. While within the Standard Model such reactions are not observable, the
possibility exists in a wide class of dynamical models for the fermion mass
generation. General arguments, based on hierarchical and democratic symmetries
are used to arrive at $t$--production rates which are detectable.
\end{abstract}
\newpage
It is well--known that within the theoretical framework of the electroweak
theory based on the gauge group $SU(2) \times U(1)$ a deeper understanding of
the observed mass spectra of the leptons and quarks is not possible. The
observations show that the masses in the various charged flavor channels
(quarks, charged leptons) are strongly dominated by the masses of the third
fermion family $m_t, m_b$
and $m_{\tau }$ respectively, a clear hierarchical pattern is observed.
It is not known whether such a mass hierarchy is present also in the neutrino
sector. In fact, it may well be that in the neutrino channel no mass
hierarchy exists, which might be the reason for the large mixing angles
indicated by the neutrino oscillation experiments \cite{fx96}.

The observed mass hierarchies and a related hierarchy of the flavor mixing
angles in the quark sector suggest that nature seems to be close to the
``rank--one limit'', in which all flavor mixing angles vanish and the mass
matrices are proportional to the rank--one matrix \cite{fri84,fri87}
\begin{equation}\label{eq1}
M_0 = {\rm const.} \left( \begin{array}{ccc}
			0 & 0 & 0 \\
			0 & 0 & 0\\
			0 & 0 & 1
			\end{array} \right).
\end{equation}

This mass matrix can also be represented, after a linear transformation of
the fermion fields, by the ``democratic'' mass matrix:
\begin{equation}\label{eq2}
\tilde{M}_0 = {\rm const.} \left( \begin{array}{ccc}
			1 & 1 & 1 \\
			1 & 1 & 1\\
			1 & 1 & 1
			\end{array} \right),
\end{equation}
which exhibits an $S(3)_L \times S(3)_R$ symmetry \cite{mdem,fp90}.

The mass generation for the members of the second and first family of 
fermions can be related to a specific breaking of the ``democratic symmetry''
\cite{fp90,fh94,fx95}.

Thus far the details of the dynamics which generate the masses and flavor
mixing
effects are unknown. They might be related, for example, to the presence of
new interactions, or to a substructure of the leptons and quarks. We shall
not speculate about those here, but rather concentrate on possible general
consequences for the physics at LEP II and
HERA. Once the masses of the fermions are generated, terms of the type
$\bar \psi_L \, \psi_R + {\rm hc}$ ($\psi$: fermion field, $L$: lefthanded,
$R$: righthanded) are introduced. These terms serve as bridges between the
lefthanded fields and the righthanded ones. In the Standard Model the mass
terms are, of course, provided by the interaction of the scalar field with
the fermions. However, in a dynamical theory of the mass generation one
expects that besides the fermion mass terms also effective interactions
between the fermions and the electroweak or QCD gauge bosons are generated,
which have a similar chiral structure as the mass terms, e. g. const.
$\bar \psi_L \, \sigma_{\mu \, \nu} \, \psi_R \, F^{\mu \, \nu}$, where
$F^{\mu \, \nu}$ stays generically for a field strength of an electroweak
gauge boson or a gluon. Of course, such terms, if present, would have to be
interpreted as signals of new interactions beyond those present in the
Standard Model.

If we consider for the quarks the ``rank--one'' limit discussed above we can
only have terms of the type
$\bar t_L \, \sigma_{\mu \, \nu} \, t_R \, F^{\mu \, \nu}$ or $\bar b_L \,
\sigma_{\mu \, \nu} \, b_R \, F^{\mu \, \nu}$ (+ h.c.). Diagonal terms like
$\bar c_L \, \sigma_{\mu \, \nu} \, c_R \, F^{\mu \, \nu}$ or flavor--changing
terms like $\bar c_L \, \sigma_{\mu \, \nu} \, t_R \, F^{\mu \, \nu}$ would
violate the chiral symmetry $SU(2)_L \times SU(2)_R$, acting on the first
two families, which is present in this limit. For example,
the flavor--diagonal terms for
the anomalous magnetic moments can be written as
\begin{equation}\label{eq3}
\frac{e}{2 \Lambda} \cdot \left( \frac{m_{i0}}{3 \Lambda} \right) \,
\bar q_L \, \sigma_{\mu \nu} \, N \, q_R \, F^{\mu \, \nu} + h. c. \, ,
\end{equation}
$\left( i = U, D, \quad m_{U0} \cong m_t, \, m_{D0} \cong m_b \right)$\\
\begin{equation}\label{eq4}
N = \left( \begin{array}{ccc}
1 & 1 & 1 \\
1 & 1 & 1 \\
1 & 1 & 1 \end{array} \right) \, .
\end{equation}
Here $\Lambda $ is a scale parameter describing the energy scale of the
underlying dynamics, which is supposedly of the order of $m_t$. After 
diagonalization it is easily seen that the anomalous magnetic moment
contributions would only be present (in the symmetry limit) for the fermions
of the third family. Flavor--violating terms which could give rise to
radiative decays like $t \rightarrow c \gamma$ or $b \rightarrow s \gamma$
do not appear.

Once this symmetry is broken and the masses of the quarks of the second
family $m_c$ and $m_s$ enter, effective transition terms between the fields
of the second and third family will in general be generated. After
diagonalization of the mass terms (i. a. after rotating away flavor changing
mass terms) a flavor mixing between the third and second family will appear.
However, there is no reason why at the same time the flavor changing
transition terms mentioned above would be rotated away. Thus we expect terms
of the type $\bar c_L \, \sigma_{\mu \, \nu} \, t_R \, F^{\mu \, \nu}$ or
$\bar s_L \, \sigma_{\mu \, \nu} \, b_R \, F^{\mu \, \nu}$ to be present. Also
in the Standard Model such effective interactions are present,
generated in higher orders of the electroweak interactions, but
in case of the $tc$--transition they are too small to be interesting for
experiment \cite{ars97}.

As an illustrative example we consider the simplest type of $S(3)_L \times
S(3)_R$--symmetry breaking discussed in Ref. \cite{fh94}. Both the
mass terms of the
charge $2/3 (U)$ and charge $- 1/3$--quarks $(D)$ have the form:
\begin{equation}\label{eq5}
M_i = M_{0i} \left( \begin{array}{ccc}
			1 & 1 & 1 \\
			1 & 1 & 1 \\
			1 & 1 & 1 + \varepsilon_i \end{array} \right)
\end{equation}
(the index $i$ stands for $U, D$ respectively).

The symmetry breaking parameter $\varepsilon $ destroys the rank one nature
of the mass matrix. Thus the mass of the second family member ($c, s$ 
respectively) is introduced. At the same time a flavor mixing angle appears,
which is proportional to the mass ratios $m_s / m_b$ and $m_c / m_t$ 
\cite{fh94}.
We also note that according to the observed mass spectrum $\varepsilon_U$ is
smaller than $\varepsilon_D$: $\varepsilon_U / \varepsilon_D \cong 0.1$.

The anomalous magnetic moments of the quarks are described by a term
proportional to
\begin{equation}\label{eq6}
\frac{e}{2 \Lambda} \cdot \left( \frac{m_{i0}}{\Lambda } \right) \bar q_L \,
\sigma_{\mu \, \nu} \, \stackrel{\sim}{N} \, q_R \cdot F^{\mu \, \nu} + h. c.
\end{equation}
where the flavor matrix $\stackrel{\sim}{N}$ is given in analogy to the mass
matrix by:
\begin{equation}\label{eq7}
\stackrel{\sim}{N}_i = \left( \begin{array}{ccc}
			  1 & 1 & 1\\
			  1 & 1 & 1\\
			  1 & 1 & 1 + \delta_i \end{array} \right)
\end{equation}
($i = U, D$ respectively).

For symmetry reasons both the mass terms and the magnetic moment terms are
given by matrices in family space where the $(3, 3)$--terms are different
from one. If the parameters $\varepsilon _i$ and $\delta _i$ were
identical, the diagonalization of the mass matrices and of the magnetic
moment matrices could be carried out using the same unitary transformation.
Thus no flavor--nondiagonal terms would arise.

Since mass terms and magnetic moments probe different properties of the
fermions, albeit they are similar in their chiral structure, there is no
reason that the symmetry breaking parameters $\varepsilon _i$ and $\delta _i$
are the same, although they are expected to be of the same order of
magnitude. Thus flavor--nondiagonal terms are expected to arise as soon as
the $S(3)_L \times S(3)_R$--symmetry is violated. In our example these terms
are proportional to $m_c / m_t$ or $m_s / m_b$ respectively:
\begin{eqnarray}\label{eq8}
{\cal L} (t \rightarrow c ) & = & {\rm const.} \cdot \frac{e}{2 \Lambda} \cdot
\left( \frac{m_t}{\Lambda} \right) \, \left( \frac{m_c}{m_t} \right) \,
\bar t_L \, \sigma_{\mu \nu} \, c_R \, F^{\mu \, \nu} + h. c. \, , \nonumber \\
{\cal L} (b \rightarrow s ) & = & {\rm const.} \cdot \frac{e}{2 \Lambda} \cdot
\left( \frac{m_b}{\Lambda} \right) \, \left( \frac{m_s}{m_b} \right) \,
\bar b_L \, \sigma_{\mu \nu} \ s_R F^{\mu \, \nu} + h.c. \, .
\end{eqnarray}
Here the const. in front depends on how far the ratios of the symmetry
breaking parameters $\delta_i / \varepsilon _i$ differ from unity.

The example discussed above shows that the $t$-- and $b$--quarks might have
properties like anomalous magnetic moments, which in principle are expected
to give rise to flavor--changing decays. How large these amplitudes are,
depends on the expected misalignment between the mass matrices and the
matrices describing the magnetic moment transitions.

>From our discussion it is evident that these misalignments are due to the
$S(3)_L \times S(3)_R$--breaking. They will not be present in the symmetry
limit. In our example they are of the order of the mixing angles, i. e. of
order $m_c / m_t$ or $m_s / m_b$. In special situations the angles describing
the misalignment can be of order $(m_c / m_t)^{1/2}$ or $(m_c / m_b)^{1/2}$;
this is the case, if the (2,2)--matrix element vanishes in the magnetic
moment matrix, if written in the hierarchy basis. In this case one finds
e. g. for the $t - c$--transition:
\begin{equation}\label{eq9}
{\cal L} ( t \rightarrow c) = {\rm const.} \cdot \frac{e}{2 \Lambda} \cdot
\left( \frac{m_c}{m_t} \right)^{1/2} \, \bar t_L \,
\sigma_{\mu \nu} \, c_R \, F^{\mu \, \nu} + h. c. \, .
\end{equation}
In order to be general we shall allow for this possibility in our discussion
below.

Both for the interaction of the $t$--quark with the photon and the
$Z$--boson we obtain anomalous vertices as follows:
\begin{eqnarray}\label{eq10}
\Delta^{tc}_Z & = & \bar c \left( i \left(C_Z + D_Z \gamma_5 \right)
\sigma^{\mu \nu} \frac{q_{\nu}}{m_t} \right) tZ_{\mu} \, , \nonumber \\
\Delta^{tc}_{\gamma} & = & \bar c \left( i \left(C_{\gamma} + D_{\gamma}
\gamma_5 \right) \sigma^{\mu \nu} \frac{q_{\nu}}{m_t} \right) tA_{\mu} \, .
\end{eqnarray}

According to our discussion above we expect for the order of magnitudes of the
vertex parameters $C, D$:
\begin{eqnarray}\label{eq11}
{\rm const.} \cdot \frac{m_c}{m_t} \cdot e & < & C_{\gamma }, \, D_{\gamma} 
 < {\rm const.} \cdot \sqrt{\frac{m_c}{m_t}} \cdot e  \nonumber \\
{\rm const.} \cdot \frac{m_c}{m_t} \cdot g_Z
& < & C_Z, \, D_Z < {\rm const.} \cdot \sqrt{\frac{m_c}{m_t}} \cdot g_Z 
\end{eqnarray}

($g_Z$: $Z$--boson coupling constant)
\begin{equation}\label{eq12}
g_z = \frac{e}{\cos\theta_{W}} \, .
\end{equation}

Thus we expect the $C, D$--parameters to be of the order of $10^{-2}$ 
down to the order of $10^{-3}$. Below we shall discuss the consequences
for $t$--decays and $t$--production in several exerpimental situations.

\begin{flushleft}
{\bf Top Decays}\\
\end{flushleft}

Besides the decay $t \rightarrow b + W$ described within the Standard
Model the decays $t \rightarrow c \, Z$, $t \rightarrow c \, \gamma$
proceed via the anomalous vertices. We find:
\begin{equation}\label{eq13}
\Gamma (t \rightarrow cZ) =
\left(|C_Z|^2 + |D_Z|^2 \right) \,
\frac{m_t}{8 \pi} \left( 1 - \frac{M^2_Z}{m^2_t} \right)
\left( 1 - \frac{1}{2} \frac{M^2_Z}{m^2_t} - \frac{1}{2} \frac{M^4_Z}{m^4_t}
\right) \, ,
\end{equation}

\begin{equation}\label{eq14}
\Gamma\left(t \rightarrow c\gamma \right) = \left(|C_{\gamma}|^2 + 
|D_{\gamma}|^2 \right) \frac{m_t}{8 \pi} \, .
\end{equation}

The standard decay is given by:
\begin{equation}\label{eq15}
\Gamma(t \to bW) = \frac{G_F}{8\pi\sqrt{2}}|V_{tb}|^2m^3_t
  \left(1-\frac{M^2_W}{m^2_t}\right)
  \left(1+\frac{M^2_W}{m^2_t}-2\frac{M^4_W}{m^4_t}\right).
\end{equation}

For the branching ratios we obtain:
\begin{equation}\label{16}
BR_Z \equiv \frac{\Gamma (t \rightarrow cZ)}{\Gamma (t \rightarrow bW)}
= \frac{\sqrt{2}}{G_F|V_{cb}|^2 m^2_t} \frac{\left( 1 - \frac{M^2_Z}{m^2_t}
\right) \left( 1 - \frac{1}{2} \frac{M^2_Z}{m^2_t} - \frac{1}{2}
\frac{M^4_Z}{m^4_t} \right)}{\left( 1 - \frac{M^2_W}{m^2_t} \right)
\left( 1 + \frac{M^2_W}{m^2_t} -2 \frac{M^4_W}{m^4_t} \right)}
\left(|C_Z|^2 + |D_Z|^2 \right) \, ,
\end{equation}
\begin{equation}\label{eq17}
BR_{\gamma} \equiv \frac{\Gamma(t \to c\gamma)}{\Gamma(t \to bW)} = 
	  \frac{\sqrt{2}}{G_F|V_{tb}|^2m^2_t}
   \frac{1}{\left(1-\frac{M^2_W}{m^2_t}\right)
  \left(1 + \frac{M^2_W}{m^2_t} - 2 \frac{M^4_W}{m^4_t} \right) }
  \left( |C_{\gamma }|^2 + |D_{\gamma}|^2 \right).
\end{equation}

For the decays $t \rightarrow c \, \gamma, \, t \rightarrow c Z$
exist the following experimental bounds \cite{cdf98}: 
\begin{eqnarray}\label{eq18}
BR (t \rightarrow q \, Z ) & < & 0.4 \, , \nonumber \\
BR (t \rightarrow q \, \gamma ) & < & 0.029 \, .
\end{eqnarray}
Thus we obtain:
\begin{eqnarray}\label{eq19}
|C_Z |^2 + |D_Z| ^2 & < & 0.16 \, ,\nonumber \\
|C_{\gamma} |^2 + | D_{\gamma }|^2 & < & 6.5 \cdot 10^{-3}.
\end{eqnarray}
For an order of magnitude estimate we take $C \cong D$ and find:
\begin{eqnarray}\label{eq20}
C_Z, \, D_Z & < & 0.3 \, ,\nonumber \\
C_{\gamma}, \, D_{\gamma } & < & 0.06 \, .
\end{eqnarray}
These bounds are significantly larger than the values estimated above,
especially those concerning the $Z$--interaction. If we take $C, D$ to be in
the upper range (see Eq. (\ref{eq11})), one finds 
$BR (t \rightarrow c \, \gamma )$
and $BR (t \rightarrow c \, Z)$ to be of the order of 1 \%. For our
subsequent discussion we shall take this branching ratio as a guideline.

\begin{flushleft}
{\bf The Production of Single t-quarks at LEP II}\\
\end{flushleft}

If the anomalous vertices given in Eq. (\ref{eq10}) are present, 
it is possible to
produce single $t$--quarks in $e^+ \, e^-$--annihilation above about
180 GeV via the reaction $e^+ \, e^- \rightarrow \gamma, \, Z \rightarrow
\bar c \, t, \, \bar t \, c$. This possibility has also been 
discussed in Ref. \cite{osy98} based on a different set of anomalous
operators. \footnote{While completing this work an additional preprint
has appeared on the same issue \protect{\cite{hep98}}.}

The total cross section for the single $t$--production is given by:
\begin{equation}\label{eq29}
\sigma_{\rm tot} \left( e^+ \, e^- \rightarrow \bar t \, c, \, \bar c \, t
\right) = \sigma_{\gamma} + \sigma_Z + \sigma_{{\rm int}} 
\end{equation}
with:
\footnote{In the following formula for $\sigma_{int}$ the imaginary parts 
of the form factors are kept for completeness only. In calculating the
cross section it was assumed that these imanginary parts vanish and
that $C$ and $D$ are of the same size. These assumptions are made for
definiteness only and should not influence the results qualitatively.}
\begin{equation}\label{eq30}
\sigma_{\gamma}=
  \frac{N_Ce^2}{16\pi m^2_t}
  \left(|C_{\gamma}|^2 + |D_{\gamma}|^2)\right) c_{\beta} ,
\end{equation}

\begin{eqnarray}\label{eq31}
\sigma_z & = &
  \frac{N_CG_FM_Z^2}{16\pi\sqrt{2}m^2_t}(1-4s_w^2+8s_w^4)
  \left(|C_Z|^2 + |D_Z|^2\right)
  \frac{s^2}{(s-M^2_Z)^2+\Gamma^2_Z M_Z^2} \, c_{\beta} ,
\end{eqnarray}

\begin{eqnarray}\label{eq32}
\sigma_{int} & = &
  \frac{N_Ce}{16\pi\sqrt{2}m^2_t}\sqrt{\frac{G_ZM^2_Z}{\sqrt{2}}}
  (1-4s_w^2)\frac{s}{(s-M^2_Z)^2+\Gamma^2_Z M_Z^2} \, c_{\beta} \nonumber\\
 & & \left[2(s-M_Z^2)\left(ReC_{\gamma}ReC_Z
  +ImC_{\gamma}ImC_Z+ReD_{\gamma}ReD_Z+ImD_{\gamma}ImD_Z\right)\right.\nonumber\\
 & & \left.-2\Gamma_Z M_Z\left(ImC_{\gamma}ReC_Z-ReC_{\gamma}ImC_Z
     +ImD_{\gamma}ReD_Z-ReD_{\gamma}ImD_Z\right)\right].
\end{eqnarray}

The following abbreviations have been used:

\begin{displaymath}
  c_{\beta}=
  \beta_{+}\beta_{-}
  \left(4\beta_{\pm}^2-\frac{2}{3}\beta^2_{+}\beta^2_{-}-2\beta^4_{\pm}
  +8\frac{m_cm_t}{s}\right) ,
\end{displaymath}
and
\begin{displaymath}
\beta^2_{+}=1-\frac{(m_t+m_c)^2}{s} \, ,\qquad
\beta^2_{-}=1-\frac{(m_t-m_c)^2}{s} \, ,\qquad
\beta^2_{\pm}=1-\frac{m_t^2-m_c^2}{s} \, .
\end{displaymath}

Numerically we find for $BR_{\gamma } = BR_Z = 1\%$ at
\begin{equation}\label{eq33}
\sqrt{s} = 190 {\rm GeV}: \sigma_{{\rm tot}} = 6.3 \cdot 10^{-2} pb 
\end{equation}
and at
\begin{equation}\label{eq34}
\sqrt{s} = 200 {\rm GeV}: \sigma_{{\rm tot}} = 13.2 \cdot 10^{-2} pb \, .
\end{equation}

In case of an integrated luminosity of 170 pb$^{-1}$/year one has 11 events
or 22 events respectively. In view of the uncertainty of the vertex
parameters $C, D$ very detailed prediction cannot be made, but our result
is encouraging. It seems not totally hopeless to look for $t$--quarks
produced singly at LEP II. Of course, if this production is observed, it
would be a clear indication towards a violation of the Standard Model
\cite{boo94}.

\begin{flushleft}
{\bf Single Production of t-quarks at HERA}\\
\end{flushleft}

It is well-known that in ep-collisions at high energy studied at
the experiments at HERA t-quarks cannot be produced at an 
observable rate if the Standard Model is valid \cite{bb88}. In our case the
reaction $ec \to et$ can take place. Note that the c-quark is
present inside the nucleon as part of the $q\overline{q}$-sea.
The differential cross section is given by:

\begin{eqnarray}\label{eq21}
\frac{d\sigma(ec \to et)}{dt} & = & 
  \frac{1}{16\pi}\sum_{V,V'=\gamma,Z} g_Vg^*_{V'} \frac{f_V}{\Lambda_{V}}
  \cdot \frac{f_{V'}}{\Lambda_{V'}} \cdot \frac{1}{D_V(t)D_{V'}(t)} \cdot
  \nonumber\\
 & &  \left\{\left(C_VC^*_{V'}+D_VD^*_{V'}\right)
	     \left(v_Vv_{V'}+a_Va_{V'}\right) 
      \left[-\left(1+\beta^4\right)t-\left(1+\beta^2\right)
	   \frac{t^2}{s}\right]
      \right. \nonumber\\
 & &  \left.+\left(C_VD^*_{V'}+D_VC^*_{V'}\right)
	     \left(v_Va_{V'}+a_Vv_{V'}\right) 
      \left[\left(1-\beta^4\right)t+\left(1-\beta^2\right)
	   \frac{t^2}{s} \right] \right\}.
\end{eqnarray}

For definiteness $\frac{f_V}{\Lambda_V}$ and $\frac{f_{V'}}{\Lambda_{V'}}$
are both fixed to $\frac{1}{m_t}$ in the following numerical examples.
Further constants are $g_{\gamma}=(-i)Q_e$, where $Q_e$ is taken to be $-e$,
and $g_Z=\left(-\frac{i}{\sqrt{2}}\right)\sqrt{\frac{G_FM^2_Z}{\sqrt{2}}}$.
The propagators in the t-channel are given by $D_{\gamma}(t)=t$ and
$D_Z(t)=t-M^2_Z$ respectively. The constants $a_V$ and $v_V$ are the standard
axialvector and vector coupling constants of $e^-$ or $e^+$ with the 
boson $V=\gamma,Z$ and finally $\beta^2=1-m^2_t/s$.

The dominant contribution comes from the $\gamma$-exchange which 
is given by:

\begin{equation}\label{eq23}
\frac{\sigma_{\gamma}(ec \to et)}{dt}= \frac{1}{8\pi} \frac{e^2}{m^2_t}
 \left(|C_{\gamma}|^2+|D_{\gamma}|^2\right)
 \left[-\frac{\left(1+\beta^4\right)}{t}
       -\frac{\left(1+\beta^2\right)}{s}\right].
\end{equation}

It becomes obvious that the cross section gets the largest where $t$
is the smallest. Note that $|t|_{min}$ is proportional to $m^2_e$ in
the limit $s \gg m^2_t$, namely

\begin{equation}\label{eq24}
|t|_{min}\simeq\frac{m^2_e(m^2_t-m^2_c)^2}{s^2}\, .
\end{equation}

The dominant contribution to the production cross section thus comes
from the phase space region near the threshold, where in the 
quark-lepton c.m.s. the incoming lepton looses all its momentum; the
energy is transferred to the quark system in order to produce the
heavy t-quark nearly at rest - in that particular frame. In the 
laboratory system both the t-quark and the lepton have no sizable
transverse momentum, and the lepton will be lost in the forward 
direction. The t-quark will decay emitting a b-quark and a W-boson.

In order to calculate the cross section for $ep \to etX$, it is 
necessary to integrate over the charm quark distribution function.
In case of HERA, where the incoming lepton has an energy of $27.6$ GeV,
the c-quark needs to have a momentum of more than 277 GeV $(x>0.338)$
such that a t-quark can be produced. Thus the cross section is
particularly large, if the c-quark distribution is as stiff as possible.

If one utilizes e.g. the charm distributions that Martin et al. have 
extracted from a global analysis \cite{mrst98} and assuming 
$BR_{\gamma}=BR_Z=1\%$, the cross section is about 
$7\cdot 10^{-4} pb$ which is too 
small to be observable at HERA. This conclusion remains the same even 
if the parameters $C$ and $D$ are allowed to saturate their bounds 
in Eq. (\ref{eq20}) because these distributions fall off quickly for 
large x. 

Alternatively we have considered charm distributions with a 
different large-x behavior. Of particular interest is the 
``intrinsic-charm''-case \cite{bro80}. The event rates have 
been calculated for the following two distribution functions 
that are based on the ``intrinsic-charm'' hypothesis \cite{bro80,gv97}:

\begin{equation}\label{eq25}
c_1(x)=\frac{1}{2}N_5x^2\left[\frac{1}{3}(1-x)(1+10x+x^2)+2x(1+x)\ln x\right],
\end{equation}

\begin{eqnarray}\label{eq26}
c_2(x) & = & \frac{1}{210}N_5x^8\left[35+1155x-1575x^2-11375x^3-2450x^4
	     +490x^5-98x^6+14x^7-x^8 \right. \nonumber\\
 & & \left.  +x\left(1443+7161x+5201x^2+
	 \left\{840+5880x+5880x^2\right\}\ln x\right)\right].
\end{eqnarray}

(Note: $N_5=36$ $(N_5=288028)$ have been used such that $c(x)$ 
is normalized to $1\%$. The distribution function $\overline{c}(x)$ is
identical to $c(x)$.)
Again making use of the assumption $BR_{\gamma}=BR_Z=1\%$ one finds:

\begin{equation}\label{eq27}
c_1(x):\qquad\sigma_{tot}=1.7\cdot 10^{-2} pb \, ,
\end{equation}

\begin{equation}\label{eq28}
c_2(x):\qquad\sigma_{tot}=3.4\cdot 10^{-2} pb \, .
\end{equation}

With an expected integrated luminosity of about 100 $pb^{-1}$/year
one could therefore hope to observe a few events of this type for
sufficient running time and favorable circumstances. 

The topology of these events is such that it could not be explained
within the Standard Model. If the $W$ decays leptonically, one would
observe a $b$-jet (or $\overline{b}$-jet) plus a high energy electron
or muon.

It is interesting to note that the H1-Collaboration has recently
reported the observation of several events with a high energy
isolated lepton \cite{h198}. Three of these events in a data sample 
of 36 $pb^{-1}$ show ``kinematic properties atypical of Standard
Model processes''. The production of single $t$- (or $\overline{t}$-)
quarks could in principle generate this type of topology and
kinematics. As a test of this hypothesis it would be mandatory
to search for those events in which the $W$ decays hadronically. 
For these events one expects to see three jets with jet momenta
that should match to give the mass and momentum of the decaying
top-quark.

Acknowledgement: We would like to thank Dr. S. Brodsky and Dr. A. Leike for
useful discussions.

\end{document}